\documentclass{article}
\usepackage{physics}
\usepackage{amssymb}
\usepackage{subfigure,graphicx}
\usepackage{arxiv}

\usepackage[utf8]{inputenc} % allow utf-8 input
\usepackage[T1]{fontenc}    % use 8-bit T1 fonts
\usepackage{hyperref}       % hyperlinks
\usepackage{url}            % simple URL typesetting
\usepackage{booktabs}       % professional-quality tables
\usepackage{amsfonts}       % blackboard math symbols
\usepackage{nicefrac}       % compact symbols for 1/2, etc.
\usepackage{microtype}      % microtypography
\usepackage{lipsum}

\usepackage{setspace}
\title{Quantum Dynamics of a One Degree-of-Freedom Hamiltonian Saddle-Node Bifurcation}

\author{
Wenyang Lyu, Shibabrat Naik, Stephen Wiggins \\
School of Mathematics, University of Bristol \\
Fry building, Woodland Road, Bristol BS8 1UG, UK
}

\begin{document}
\maketitle

\begin{abstract}
In this paper, we study the quantum dynamics of a one degree-of-freedom (DOF) Hamiltonian that is a normal form for a saddle node bifurcation of equilibrium points in phase space. The Hamiltonian has the form of the sum of kinetic energy and potential energy. The bifurcation parameter is in the potential energy function and its effect on the potential energy is to vary the depth of the potential well. The main focus is to evaluate the effect of the depth of the well on the quantum dynamics.  This evaluation is carried out through the computation of energy eigenvalues and eigenvectors of the time-independent Schr{\"o}dinger equations, expectation values and position uncertainties for position coordinate, and Wigner functions.
\end{abstract}

% Keywords

\tableofcontents

\section{Introduction}
Recently, Refs.~\cite{garcia2020tilting,lyu2020role} studied the effect of depth and flatness of the potential energy function on a classical Hamiltonian and showed the influence of depth on reaction dynamics. The authors defined the depth of the potential energy as the potential energy difference between the saddle equilibrium point and the centre equilibrium point. In this work, we follow their formulation and study the effect of the depth of the potential energy on the quantum dynamics of the one DOF Hamiltonian which undergoes saddle node bifurcation. The Hamiltonian has a saddle equilibrium point at the origin and a centre equilibrium point on the right of the saddle with negative total energy in phase space. The saddle node bifurcation occurs as the depth of the potential energy (to be referred to as well-depth) is decreased and the center equilibrium point collides with the saddle equilibrium point. The classical dynamics can be understood qualitatively and quantitatively~\cite{lyu2020role} since the system is integrable and trajectories lie on the level curves of the Hamiltonian. Furthermore, reaction defined as the crossing of the saddle equilibrium point is only possible when the energy is positive (since the energy of the saddle equilibrium point is zero). However, the complete picture of reaction dynamics needs studying the quantum dynamics of this one DOF system en route to the saddle-node bifurcation. When the total energy is between the total energy of the saddle equilibrium point and the centre equilibrium point, the solution of the time-independent Schr{\"o}dinger equation consists of scattering state~\cite{ferreira2014global}, tunnelling and bound state regions. Our goal is to understand the effect of the well-depth on the quantum dynamics as a means to study the correspondence between classical and quantum dynamics. 

% Classical reaction dynamics can be understand by looking at the phase space structures of the Hamiltonian. However, the quantum dynamics is are complicated. 
% Thus, studying the effect of the well-depth  on the quantum dynamics improves our understanding on the quantum reaction dynamics and helps us identify the correspondence between classical and quantum mechanics. 

Quantum mechanical aspects of saddle node bifurcation of periodic orbits  have been studied by other authors. For example, Refs. ~\cite{borondo1995quantum,borondo1996saddle} studied the saddle node bifurcation of periodic orbits in the LiNC/LiCN system. The authors computed several quantities and study the effect of the saddle node bifurcation in both the classical and the quantum mechanics. Refs.~\cite{uwano1998geometric,uwano1999quantum} studied the saddle node bifurcation of periodic trajectories in a two DOF Hamiltonian system. The authors showed that for certain parameter values, a degeneracy of energy levels occurs and concluded that the degeneracy is a quantum analogue of the saddle-node bifurcation of periodic trajectories in classical mechanics.

% Unlike classical mechanics, quantum mechanics is built on Schr{\"o}dinger equation where the classical Hamiltonian is replaced by a Hamiltonian operator~\cite{wiggins_2020}. Even for the one dimensional system, only few analytical solutions of the Schr{\"o}dinger equation are available, for e.g., the harmonic oscillator~\cite{wiggins_2020}, the Morse oscillator~\cite{Dahl1988}, and the inverted harmonic oscillator~\cite{barton1986quantum}.The Schr{\"o}dinger equation is a linear partial differential equation and different boundary conditions can lead to different solutions, for e.g., the harmonic oscillator~\cite{Consortini1976}, the inverted harmonic oscillator~\cite{roy2015quantum}. Therefore, finding an appropriate method for solving the Schr{\"o}dinger equation is always a task.  When the Hamiltonian does not depend on time, the Schr{\"o}dinger equation can be simplified to its time-independent version. 

Analytical solutions of the time-independent Schr{\"o}dinger equation are known for some special one DOF systems, for example, the harmonic oscillator~\cite{griffiths2018introduction,wiggins_2020}, the Morse oscillator~\cite{Dahl1988}, and the inverted harmonic oscillator~\cite{barton1986quantum}. The choice of the boundary condition and an appropriate numerical method are critical steps in solving the boundary value problem for a linear partial differential equation which is the eigenvalue problem associated with solving the time-independent Schr{\"o}dinger equation; for example, the harmonic oscillator~\cite{Consortini1976}, the inverted harmonic oscillator~\cite{roy2015quantum}. In the Schr{\"o}dinger's formulation of quantum mechanics, solving the time-independent Schr{\"o}dinger equation is treated as an eigenvalue problem where the eigenvalue is referred to as the energy  and the eigenvector is called the energy eigenfunction. Since the energy eigenfunction is a configuration space quantity, we study the correspondence between classical and quantum dynamics using the phase space quantity called the Wigner function~\cite{Wigner1932, Hillery1984121, Kim_Wigner1990}. 

% Solving the time-independent Schr{\"o}dinger equation constitutes solving a boundary value problem (BVP) for a linear partial differential equation, and thus the solutions are only unique for a given boundary condition. The choice of the boundary condition and an appropriate numerical method are critical steps in solving the BVP; 

Formally, the Wigner function is defined as the Weyl transform~\cite{weyl1931group} of the density operator and is a joint distribution function of the position and the momentum coordinate. However, the Wigner function can contain negative values and therefore is not a simple probability distribution. This is due to the fact that we can not find a non negative probability distribution in phase space~\cite{Hillery1984121}, in contrast to classical mechanics. Hence, the Wigner function is usually called a quasi probability distribution in phase space. Despite having negative values, the Wigner function is a normalised density function and it can be used to calculate expectation values of functions of the position and the momentum coordinate. References about the Wigner function can be found in~\cite{Hillery1984121,lee1995theory,case2008wigner,campos1998correlation,styer2002nine,belloni2004wigner}.

Although the potential energy function is a configuration space concept, the geometry of the potential energy function is important in both classical and quantum mechanics in phase space and is closely related to reaction dynamics~\cite{waalkens_wigners_2008}. For example, the critical points of the potential energy function is directly related to equilibrium points in phase space. For systems of 2 or more DOF, the dynamical manifestation of saddle type critical points of the PES are normally hyperbolic invariant manifold which is the fundamental object in constructing a dividing surface in the transition state theory~\cite{wiggins2016role,agaoglou_chemical_2019}. 
%Also, the existence of the valley ridge inflection point on the potential energy function can lead to mutiple products which are important in reactions that involve post-transition state bifurcations~\cite{hare2017post}. 
Furthermore, it is now understood that the depth of a potential well affects the product ratios, formation of specific isomers, and identification of the intrinsic reaction coordinate in chemical reactions~\cite{koseki_intrinsic_1989,nummela_nonstatistical_2002}. 
% and a trajectory enters a deeper well is likely to take more time to escape from the well. 
We study the influence of the well-depth on the classical reaction dynamics and its correspondence with the quantum dynamics for a Hamiltonian that shows saddle node bifurcation.
% However, the influence of depth of the potential energy function on both classical and quantum dynamics of chemical reactions have not been investigated using a quantitative approach. 

This article is structured as follows. In section 2 we briefly discuss the classical dynamics of a Hamiltonian that the equilibrium points of it undergo saddle node bifurcation and the formulation of the depth of the potential energy function. We then introduce the quantum version of the Hamiltonian and its time-independent Schr{\"o}dinger equation. In section 3 we discuss the influence of depth of the potential energy function on the quantum dynamics. Both qualitative and quantitative results are present including the energy eigenvalues and eigenvectors of the time-independent Schr{\"o}dinger equations, expectation values and position uncertainties for position coordinate, and Wigner functions. The conclusions and outlook are present in section 4.
 
%%%%%%%%%%%%%%%%%%%%%%%%%%%%%%%%%%%%%%%%%%
\section{Classical and quantum dynamics}
\label{sec:headings}
We consider the normal form for the one DOF Hamiltonian that the equilibrium points of it undergo saddle node bifurcation in phase space~\cite{Wiggins2017book}. This is given by
\begin{equation}
\mathcal{H}(x,p_x) = \mathcal{T}(p_x) + \mathcal{V}(x) = \frac{1}{2} \, p_x^2 - \sqrt{\mu} \, x^2 + \frac{\alpha}{3} \, x^3 \;,
\label{eqn:ham1dof_snbif}
\end{equation}
where parameter $\mu \geq 0$ controls the location of one of the equilibrium points relative to another and  $\alpha > 0$ is the well-depth parameter and denotes the strength of the nonlinear terms in the potential energy. In this system, we define the \emph{reaction} as the change in sign of the $x$-coordinate, and in particular, we specify reaction to be the event when a trajectory goes from $x > 0$ to $x < 0$. 

The Hamiltonian vector field given by Eqn.~\eqref{eqn:ham1dof_snbif} has two equilibrium points (also referred to as critical points of the potential energy function and obtained by setting the gradient of the potential energy to zero) at $\mathbf{x}_1^e= (0,0)$ and $\mathbf{x}_2^e=(2\sqrt{\mu}/\alpha,0)$. Linear stability of the equilibria is given by the eigenvalues of the Jacobian of the vector field (or the Hessian of the potential energy function). This gives that $\mathbf{x}_1^e$ is a saddle and $\mathbf{x}_2^e$ is a centre equilibrium point. The energy of these equilibrium points are 
\begin{align}
\mathcal{H}(\mathbf{x}_1^e) = 0 \;,\quad \mathcal{H}(\mathbf{x}_2^e) = - \dfrac{4\mu^{3/2}}{3\alpha^2}.    
\end{align}

\begin{figure}[!ht]
	\centering
	\includegraphics[width=0.45\linewidth]{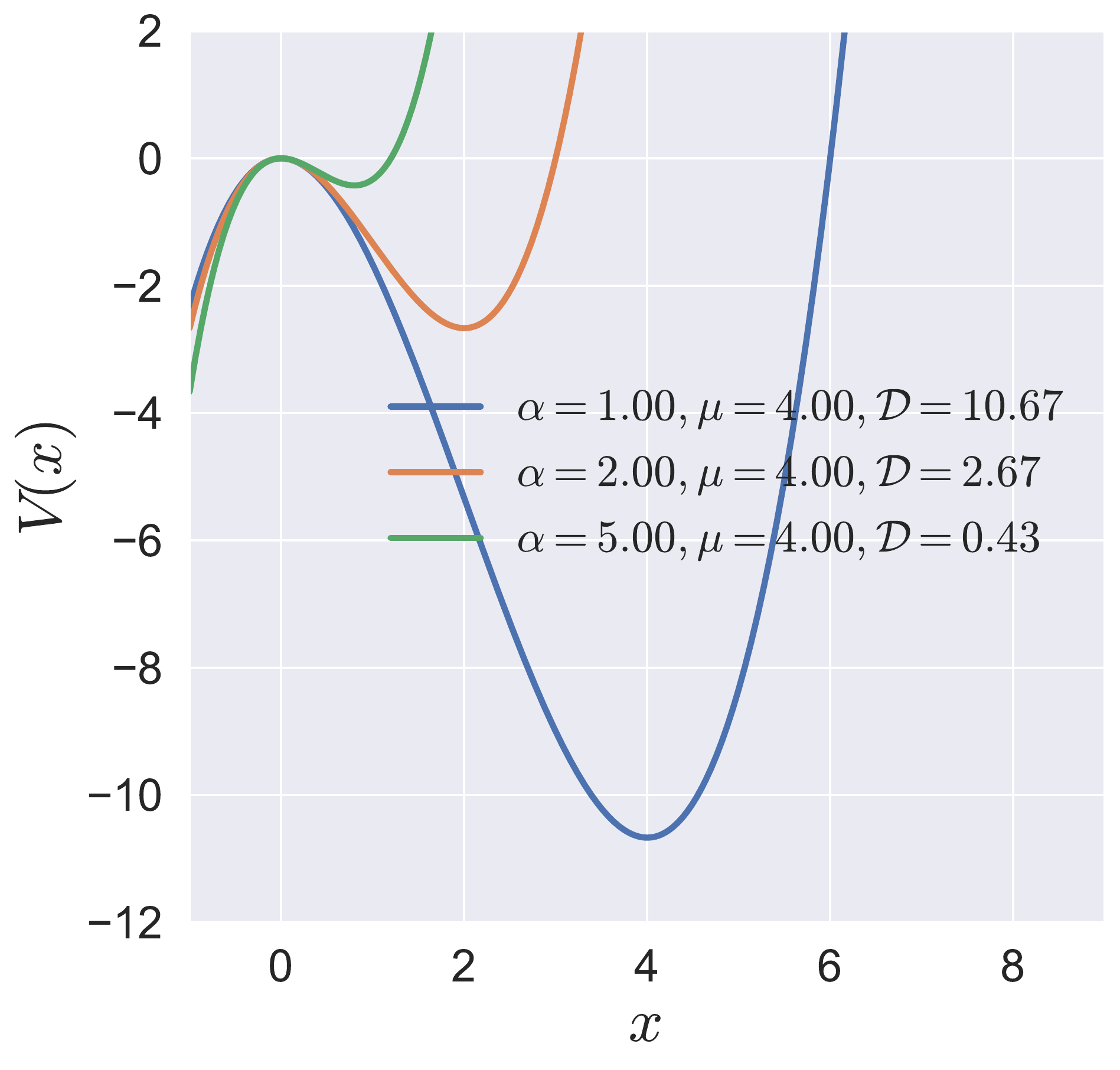}
	\caption{Potential energy function $V(x)$ for $\alpha = 1, 2, 5, \mu = 4$.}
	\label{fig:quantum_sd1DOF_pes}
\end{figure}

The one DOF saddle-node Hamiltonian is integrable for all parameter values and trajectories lie on the isoenergetic contours given by the Hamiltonian~\eqref{eqn:ham1dof_snbif}. 
The phase space trajectories with total energy $e$ can be classified as reactive (with positive total energy $e > 0$) or non-reactive (with negative total energy $e < 0$) by simply checking if the isoenergetic contour ($\mathcal{H}(x,p_x) = e$) crosses $x = 0$.
We follow the formulation in Ref.~\cite{lyu2020role} which defined the depth $\mathcal{D}$ of the potential energy function as the the difference between the potential energy of the saddle equilibrium point and the potential energy of the centre equilibrium point:
\begin{equation}
    \mathcal{D} = V(\mathbf{x}_1^e) - V(\mathbf{x}_2^e) = \dfrac{4\mu^{3/2}}{3\alpha^2}.
\end{equation}
The classical reaction dynamics in the one DOF system has been studied in Ref.~\cite{garcia2020tilting,lyu2020role} and we will not repeat it here. We present the one dimensional potential energy function for $\alpha = 1, 2, 5, \mu = 4$ in Fig.~\ref{fig:quantum_sd1DOF_pes}.
We can see from the figure that for a fixed value of $\mu$, larger value of $\alpha$ means that we have a deeper potential well.

% \textbf{Phase space view of the classical dynamics.} \sn{Working on this section}

\begin{figure}[!ht]
	\centering
	\includegraphics[width=0.65\linewidth]{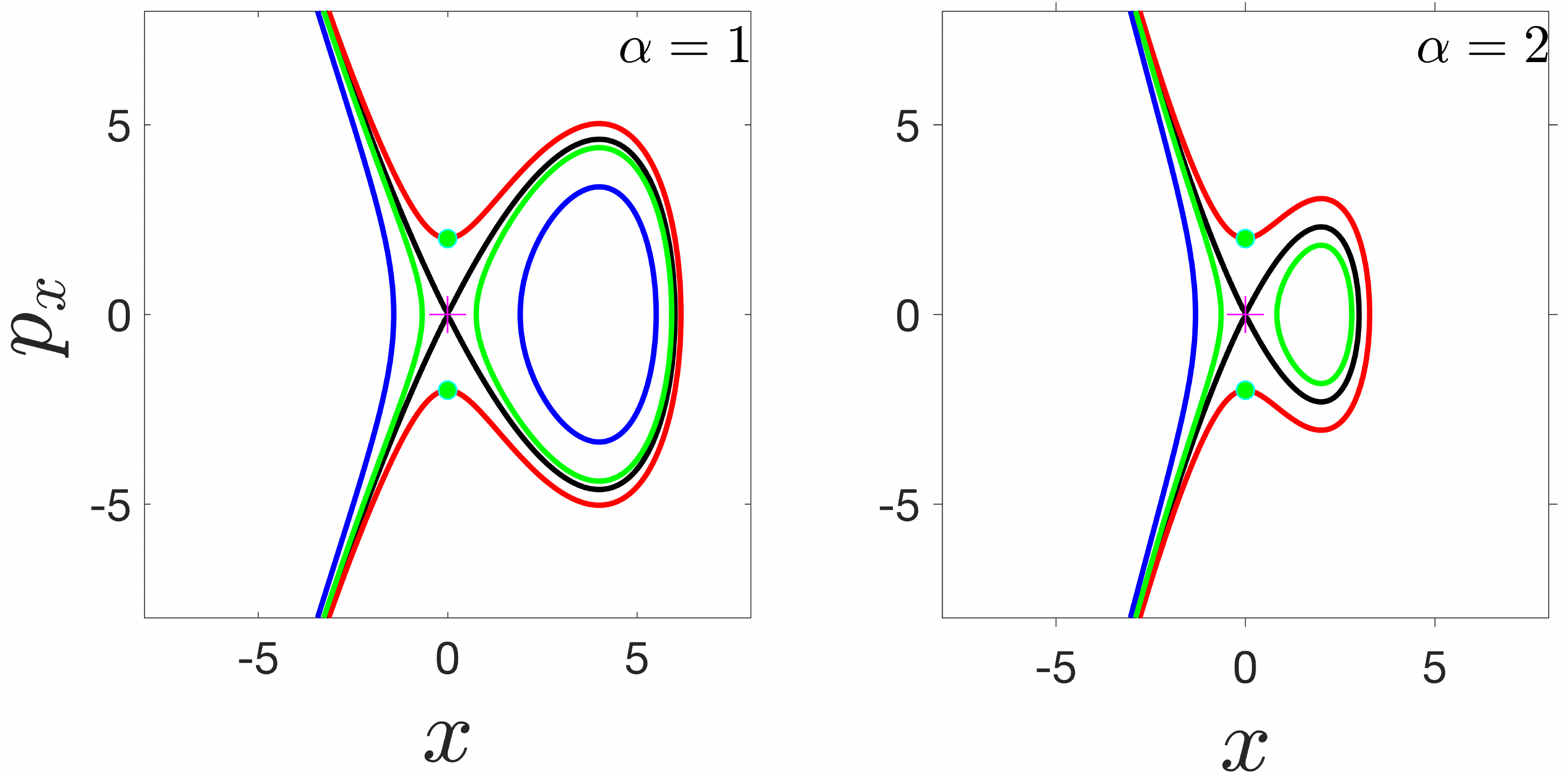}
	\caption{Phase space view of the classical dynamics showing the change in the geometry of the reactive trajectories (shown in red, $e = 2$) and non-reactive trajectories (shown in blue $e = -5$ and green $e = -2$) for different values of the parameter $\alpha$ and fixed $\mu = 4$.}
	\label{fig:saddlenode1dof_mu4}
\end{figure}

The geometry of the phase space structures explain the mechanism behind the reactive and non-reactive trajectories~\cite{uzer2002geometry}. The phase space structure in the bottleneck is the saddle equilibrium point at the origin which is also a normally hyperbolic invariant manifold (NHIM)~\cite{wiggins2016role}. We note here that only for a one dimensional potential energy function the NHIM (shown as red plus in Fig.~\ref{fig:saddlenode1dof_mu4}) does not change with total energy of the system, and in general the NHIM depends on the total energy. Next, the trajectories can be separated by constructing a dividing surface at total energy $\mathcal{H}(x,p_x) = e$. These are the points (shown as cyan dots) on the isoenergetic contour (shown as red curve) above the energy of the saddle equilibrium point in the Fig.~\ref{fig:saddlenode1dof_mu4}. The reaction dynamics at different total energies can now be classified as the reactive trajectories shown as red and black curves, or non-reactive trajectories shown as green or blue curves, respectively. As the parameters of the potential energy surface are varied, the geometry of the reactive trajectories in the phase space can be inferred from the isoenergetic contours as shown in the Fig.~\ref{fig:saddlenode1dof_mu4}. 
% We also note that going from $\alpha = 1$ to $\alpha = 2$, the phase space volume inside the isoenergetic curves for $e \geq 0$ (bounded by the red curve and to the right of the origin) decreases. This implies that we need to enforce equal density when initializing reactant volume for calculations comparing different parameter values. We will return to this system after developing the formulation for depth and flatness. 

As the Hamiltonian~\eqref{eqn:ham1dof_snbif} is time-independent, the quantum dynamics is governed by the time-independent Schr\"{o}dinger equation with the Hamiltonian operator
\begin{align}
\hat{\mathcal{H}} \psi(x) = & E \psi(x) \label{eqn:time_inde_Schrodinger_1d}    \\
\text{where},
\hat{\mathcal{H}}(\hat{x},\hat{p_x}) = \hat{\mathcal{T}}(\hat{p_x}) + \hat{\mathcal{V}}(\hat{x}) = & \frac{1}{2} \, \hat{p_x}^2 - \sqrt{\mu} \, \hat{x}^2 + \frac{\alpha}{3} \, \hat{x}^3 \label{eqn:ham_operator1dof_snbif} \;,
\end{align}
where $\hat{\mathcal{T}}$ is the kinetic energy operator, $\hat{\mathcal{V}}$ is the potential energy operator, $E$ is the energy level (or eigenvalue of the Hamiltonian operator) and $\psi(x)$ is its associated energy eigenfunction. The solution of the time-independent Schr\"{o}dinger equation is of the form: 
\begin{equation}
\hat{\mathcal{H}} \psi_n(x) = E_n \psi_n(x),\quad n = 0, 1, 2, ...
\end{equation}
where $E_n$ is the $n$-th energy level and $\psi_n(x)$ is the $n$-th energy eigenfunction. Although the classical dynamics is simple, the quantum dynamics of the 1 DOF system is not so straightforward. The global solution of the Schr{\"o}dinger equation can be found using power series expansion and is studied in for e.g. ~\cite{ferreira2014global}. The global solution consists of bound state, scattering state and tunnelling behaviours. For our purpose, it is enough to take a finite interval around the saddle and the centre equilibrium points and study the effect of depth on the quantum dynamics within this interval.

\section{Results}
In this section we present the qualitative and quantitative effect of the well-depth on quantum dynamics by calculating the energy eigenvalues and eigenfunctions of the time-independent Schr{\"o}dinger equation, mean position and position uncertainty of $n$-th (for $n=0,1,2,3,4$) excited state, and the Wigner function.

% We choose this interval is because the bound state behaviour presents in this domain, for all the values of parameters we considered. The value of $\mu$ is taken to be $4$ and we vary $\alpha$ to change the value of $\mathcal{D}$.
\subsection{Quantum states}
\begin{figure}[!ht]
	\centering
	%\subfigure[$\alpha=1$, FGH]{\includegraphics[width=0.33\textwidth]{figures/quantum_sd1DOF_eigenvec_FGH_mu4_alpha1.pdf}}
    %\subfigure[$\alpha=2$, FGH]{\includegraphics[width=0.33\textwidth]{figures/quantum_sd1DOF_eigenvec_FGH_mu4_alpha2.pdf}}
    %\subfigure[$\alpha=5$, FGH]{\includegraphics[width=0.33\textwidth]{figures/quantum_sd1DOF_eigenvec_FGH_mu4_alpha5.pdf}}
    \subfigure[$\mathcal{D}=10.67$]{\includegraphics[width=0.32\textwidth]{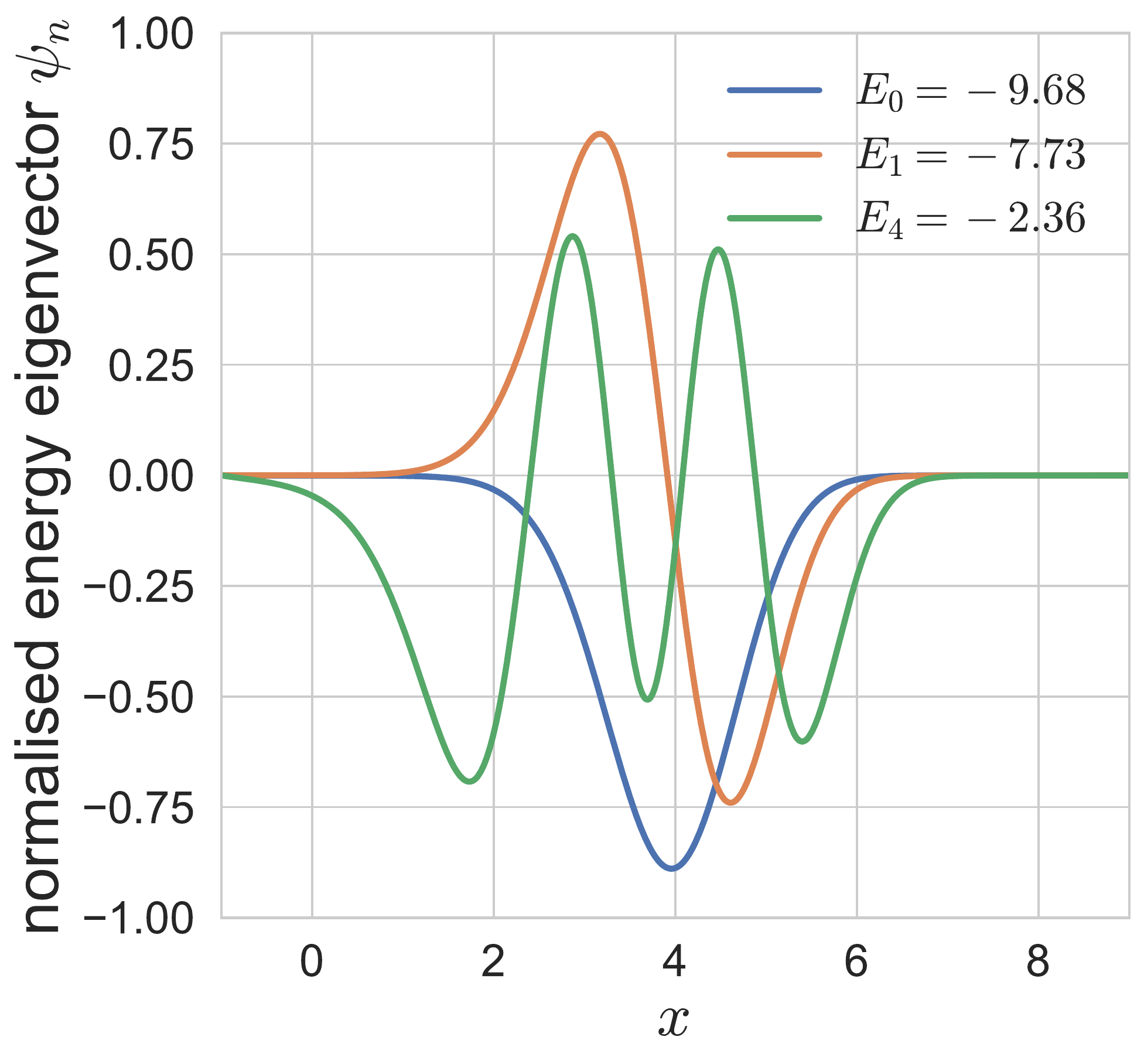}}
    \subfigure[$\mathcal{D}=2.67$]{\includegraphics[width=0.32\textwidth]{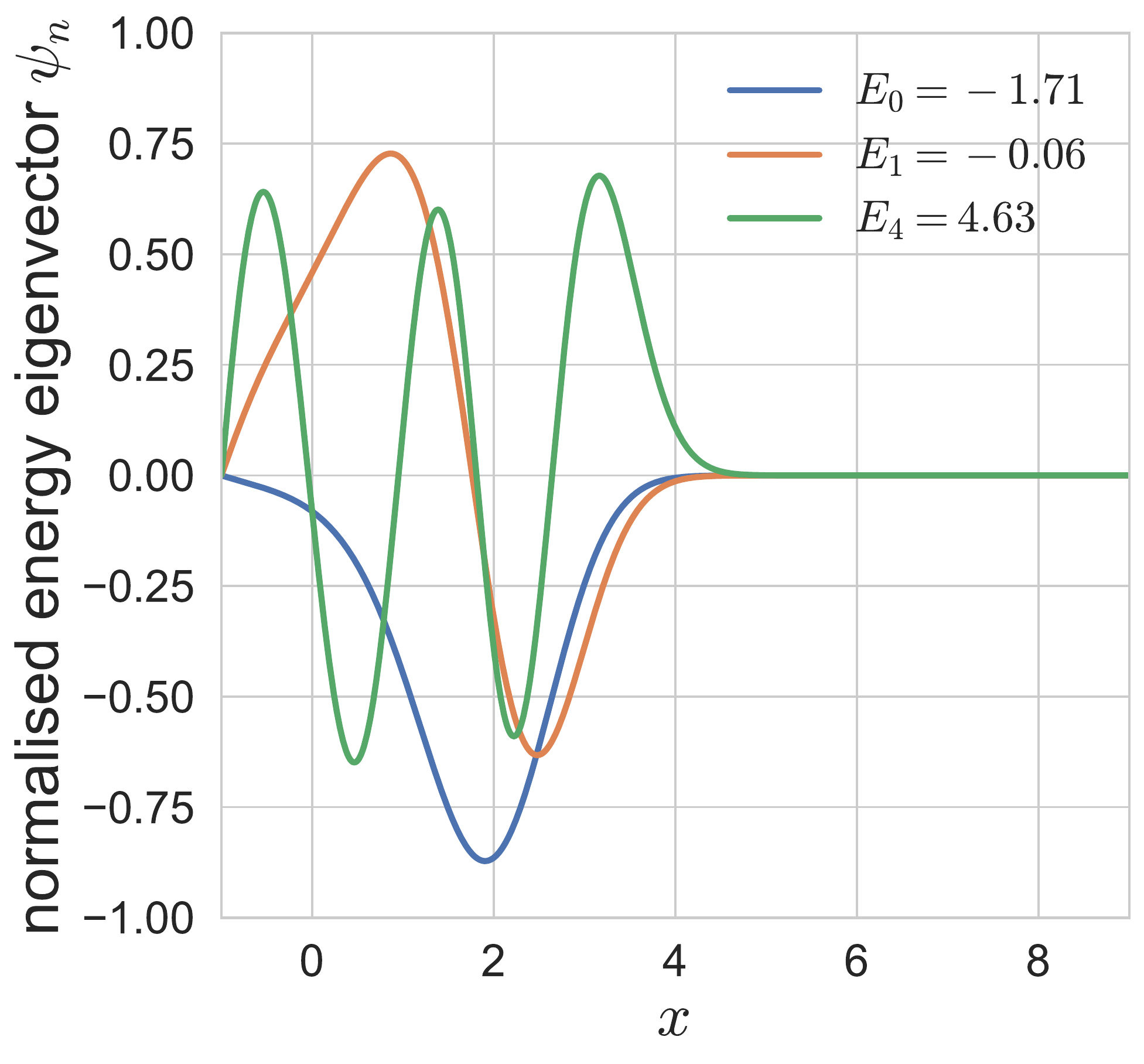}}
    \subfigure[$\mathcal{D}=0.43$]{\includegraphics[width=0.32\textwidth]{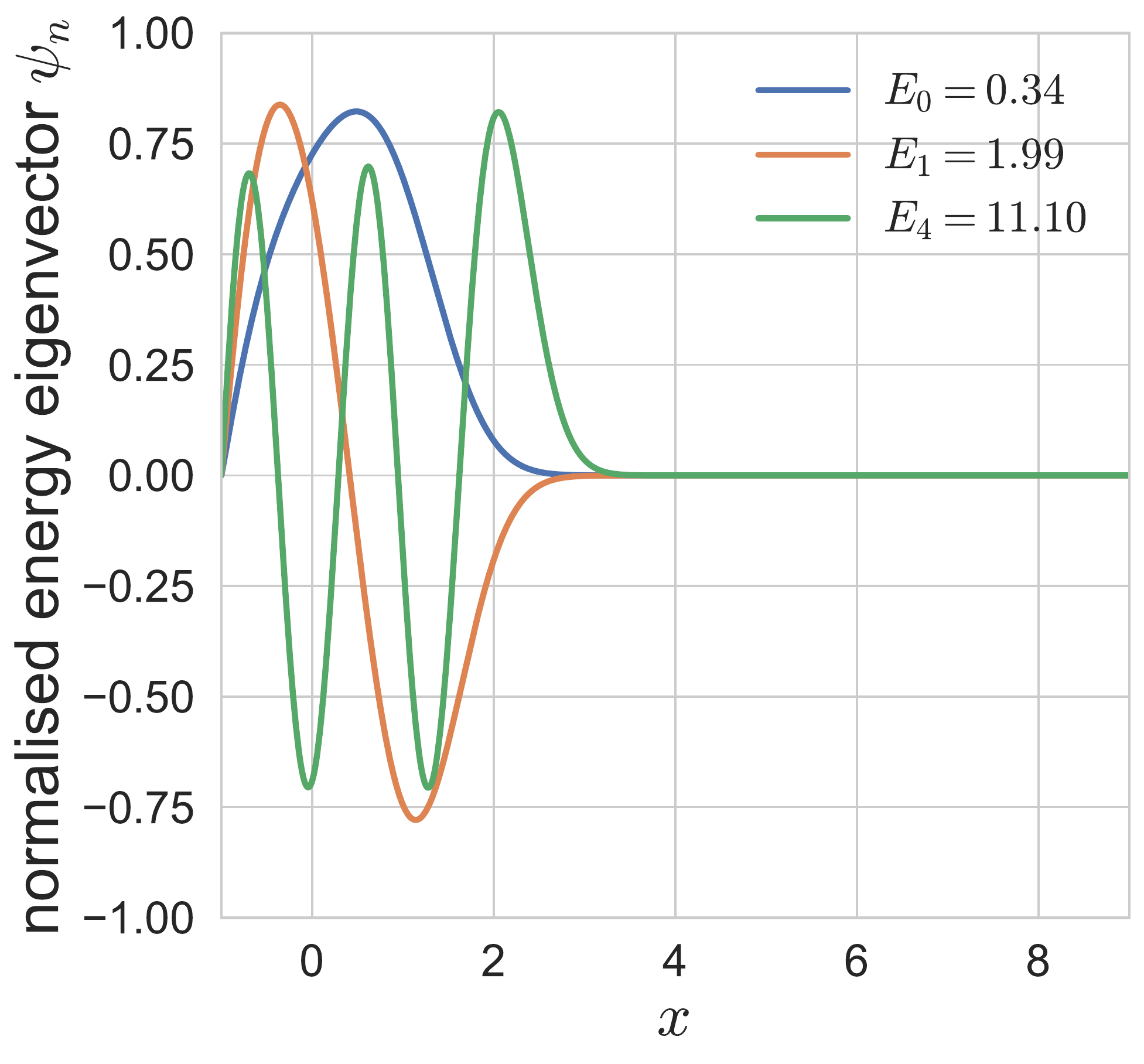}}
	\caption{Energy eigenvalues $E_n$ and energy eigenfunctions $\psi_n$ for $n=0,1,4$ energy state obtained using the FD method. The coordinate space is chosen as $[-1, 9]$, number of grid points $N=599$, $m=1$, $\alpha=1, 2, 5, \mu=4$ and $\hslash=1$.}
	\label{fig:quantum_sd1DOF_eigenvec}
\end{figure}

We compute the energy eigenvalue $E_n$ and energy eigenfunctions $\psi_n$ for the time-independent Schr\"{o}dinger equation (Eqn.~\eqref{eqn:ham_operator1dof_snbif}) on the domain $[-1, 9]$. This interval shows bound state behaviour for all values of parameters considered in this study. We solve the eigenvalue problem by discretising the Hamiltonian operator in the time-independent Schr{\"o}dinger equation using the Finite Difference (FD) method. We keep the $\mu = 4$ fixed and vary $\alpha$ to change the well-depth, $\mathcal{D}$. We show the energy eigenvalues $E_n$ and energy eigenfunctions $\psi_n$ for $n = 0,1,4$ energy state for the one DOF system using the FD method in Fig.~\ref{fig:quantum_sd1DOF_eigenvec} for three fixed values of the well-depth. 

We observe that for small $n$, the eigenfunction behaves like a bound state solution, and for large $n$, eigenfunction behaves like a plane wave on the left and approaches 0 on the right which is a sign of a scattering state solution. As $\mathcal{D}$ decreases, bound state behaviour of the energy eigenfunctions disappears. This makes sense as the time-independent Schr{\"o}dinger equation has scattering state solution when the depth of the potential energy surface is zero which corresponds to $\mathcal{D} = 0$ in this case~\cite{ferreira2014global}.

\begin{figure}[!ht]
	\centering
    \subfigure[]{\includegraphics[width=0.32\textwidth]{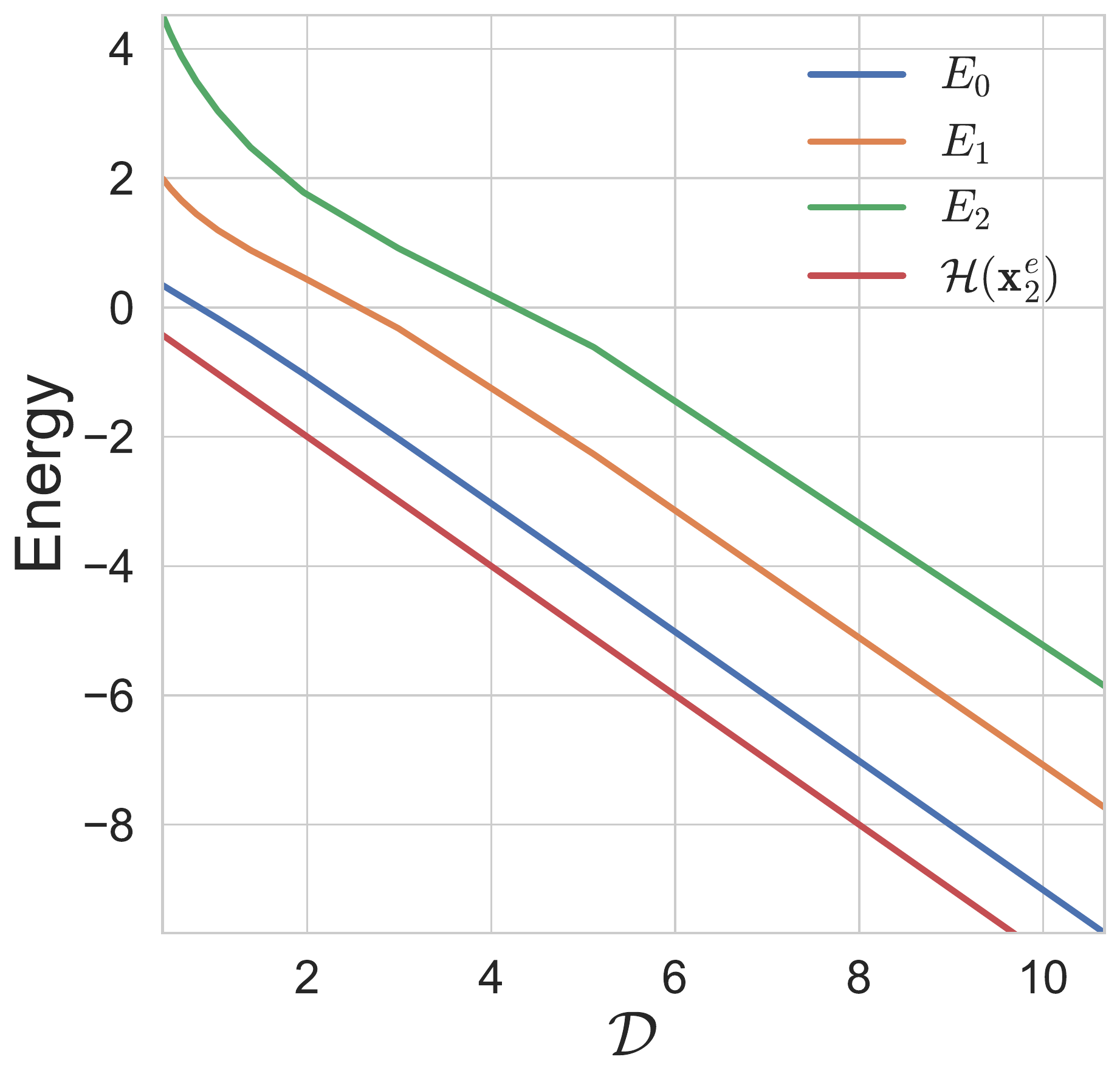}}
    \subfigure[]{\includegraphics[width=0.32\textwidth]{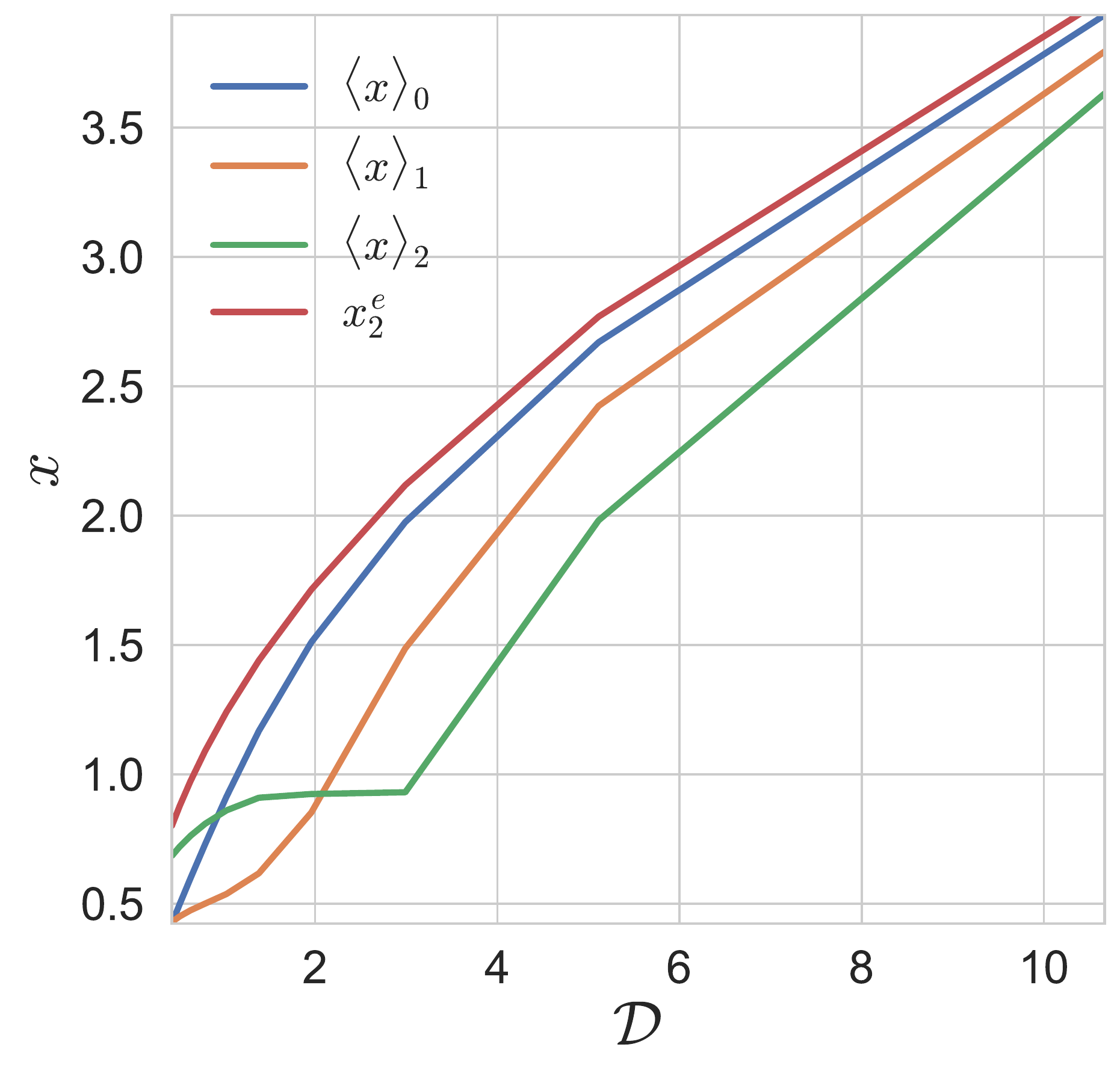}}
    \subfigure[]{\includegraphics[width=0.32\textwidth]{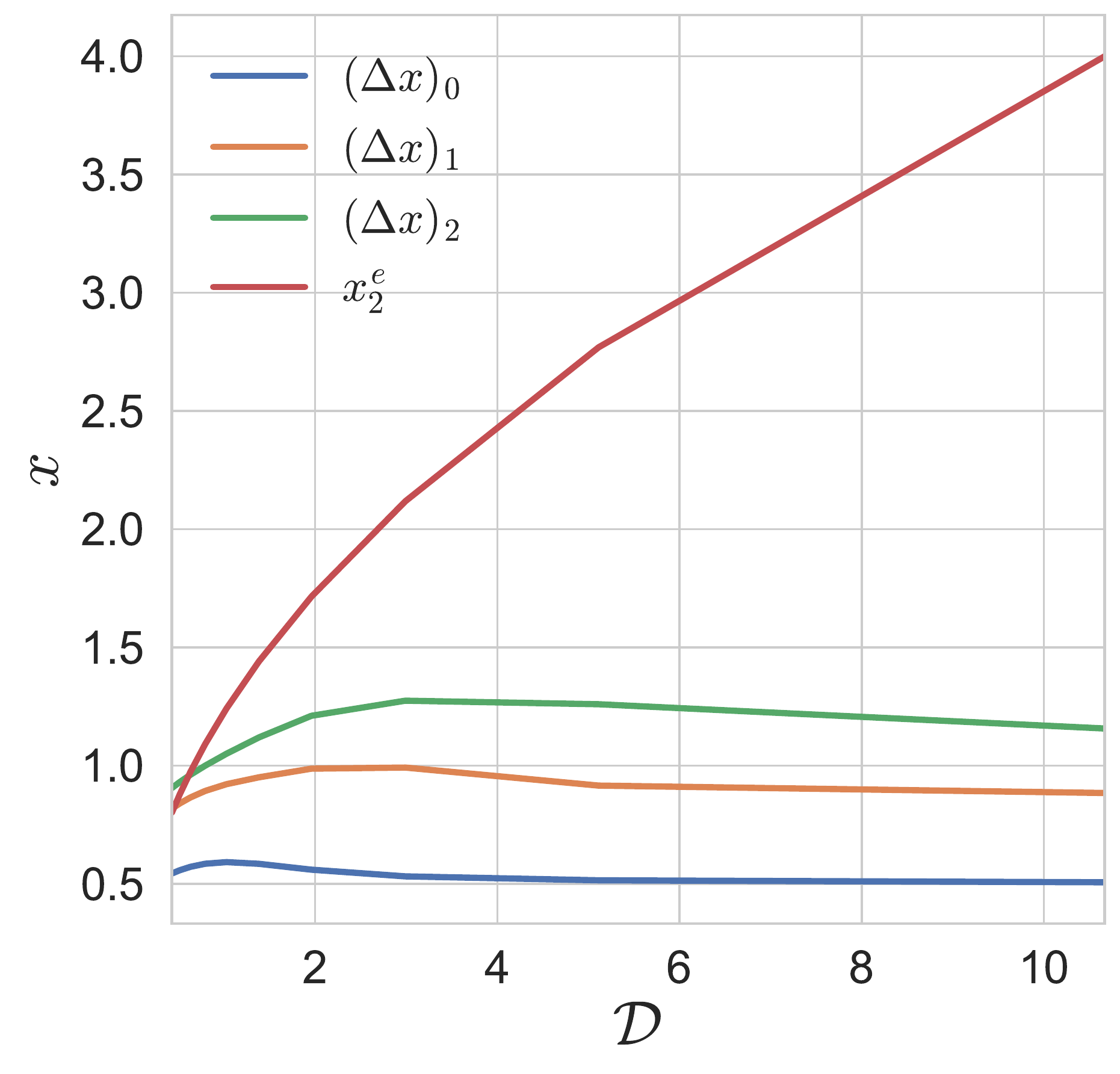}}
	\caption{(a) Energy eigenvalues $E_n$, (b) mean position of $n$-th excited state $\langle x\rangle_{n}$, and (c) position uncertainty of eigenfunctions of the $n$-th energy state for $n = 0,1,2$ energy state against well-depth, $\mathcal{D}$. The energy of the centre equilibrium point $\mathcal{H}(\mathbf{x}_2^e)$ in (a), the $x$ coordinate of the centre equilibrium point $x_2^e$ in (b), (c) are also shown for comparison. The coordinate space is chosen as $[-1, 9]$, number of grid points $N=599$, $m=1$, $\alpha \in [1, 5], \mu=4$ and $\hslash=1$.}
	\label{fig:quantum_sd1DOF_evalue&mean_pos_dep}
\end{figure}

The mean position of $n$-th excited state $\langle x\rangle_{n}$ is defined as 
\begin{equation}
    \langle x\rangle_{n} = \int \psi_n^*(x) x \psi_n(x) dx,
\end{equation}
and we can use it to calculate the position uncertainty of $n$-th excited state
\begin{equation}
    (\Delta x)_n = \sqrt{\langle x^2 \rangle_{n} - (\langle x \rangle_n)^2}.
\end{equation}
In Fig.~\ref{fig:quantum_sd1DOF_evalue&mean_pos_dep}, we show the energy eigenvalues $E_n$, mean position of $n$-th excited state $\langle x\rangle_{n}$, and position uncertainty for $n = 0,1,2$ excited state $( \Delta x)_n$ against $\mathcal{D}$. The energy of the centre equilibrium point $\mathcal{H}(\mathbf{x}_2^e)$ and $x$ coordinate of the centre equilibrium point $x_2^e=2\sqrt{\mu}/\alpha$ are also shown. We vary the well-depth by varying $\alpha$ over the interval $[1,5]$.

% 
% We can see from Fig.~\ref{fig:quantum_sd1DOF_evalue&mean_pos_dep} (a) that for the same value of $\mathcal{D}$, compare with the energy eigenvalues, the energy of the centre equilibrium point is the smallest. The ground state energy eigenvalue is smaller than the first excited state energy eigenvalue and is smaller than the second excited state energy eigenvalue. 

For the well-depth and domain considered in the numerical computation for the Fig.~\ref{fig:quantum_sd1DOF_evalue&mean_pos_dep}, we see the energy of the ground state and excited states are above the energy of the well. 
As the well-depth increases, energy eigenvalues of all energy states decrease. We observe a linear relationship between $\mathcal{D}$ and $E_n$ which can be explained using Eqn.~\eqref{eqn::quantum_pes_harm+perturb}. By comparing the mean position of all three states in Fig.~\ref{fig:quantum_sd1DOF_evalue&mean_pos_dep} (b), we see that for the same value of well-depth, $\mathcal{D}$, the mean position of a given energy state lies between the saddle and the center equilibria. As $n$ increases, that is for higher excited states, the mean position moves closer to the saddle equilibrium point and away from the centre equilibrium point. As the well-depth, $\mathcal{D}$, decreases, the $x$-coordinate of the centre equilibrium point moves closer to the saddle origin and the mean position of $n$-th excited state energy eigenfunction also moves closer to the saddle at the origin. Fig.~\ref{fig:quantum_sd1DOF_evalue&mean_pos_dep} (c) shows that for the same value of $\mathcal{D}$, the position uncertainty of the ground state is smaller than the position uncertainty of the first excited state whose position uncertainty is smaller than the same of the second excited state. As the value of $\mathcal{D}$ increases, we observe the position certainties for all three states becomes constant. 

%The energy eigenvalues $E_n$ again $\mathcal{D}$ of the 1 DOF system are shown in Fig.~\ref{fig:quantum_sd1DOF_evalue_alpha&dep&flat}. The range of $\alpha$ is between $1$ and $5$.

%\begin{figure}[!ht]
	%\centering
    %\subfigure[]{\includegraphics[width=0.33\textwidth]{figures/quantum_sd1DOF_evalue_alpha.pdf}}
    %\subfigure[]{\includegraphics[width=0.33\textwidth]{figures/quantum_sd1DOF_evalue_depth.pdf}}
    %\subfigure[]{\includegraphics[width=0.33\textwidth]{figures/quantum_sd1DOF_evalue_flat.pdf}}
	%\caption{\textbf{Energy eigenvalues $E_n$ against $\mathcal{D}$.} The coordinate space is chosen as $[-1, 9]$, number of grid points $N=599$, $m=1$, $\alpha \in [1, 5], \mu=4$ and $\hslash=1$.}
	%\label{fig:quantum_sd1DOF_evalue_alpha&dep&flat}
%\end{figure}

%The mean position of eigenvectors $\langle\psi_{n}(x)\rangle$ again $\mathcal{D}$ of the 1 DOF system are shown in Fig.~\ref{fig:quantum_sd1DOF_mean_pos_alpha&dep&flat}. The range of $\alpha$ is between $1$ and $5$.

%\begin{figure}[!ht]
	%\centering
    %\subfigure[]{\includegraphics[width=0.33\textwidth]{figures/quantum_sd1DOF_mean_pos_alpha.pdf}}
    %\subfigure[]{\includegraphics[width=0.33\textwidth]{figures/quantum_sd1DOF_mean_pos_dep.pdf}}
    %\subfigure[]{\includegraphics[width=0.33\textwidth]{figures/quantum_sd1DOF_mean_pos_flat.pdf}}
	%\caption{\textbf{Mean position of eigenvectors $\langle\psi_{n}(x)\rangle$ against $\mathcal{D}$.} The coordinate space is chosen as $[-1, 9]$, number of grid points $N=599$, $m=1$, $\alpha \in [1, 5], \mu=4$ and $\hslash=1$.}
	%\label{fig:quantum_sd1DOF_mean_pos_alpha&dep&flat}
%\end{figure}

\subsection{Wigner function}

We first compute the Wigner function for the one DOF system and then calculate an integral based on the Wigner function to quantify the distribution in the phase space. 

% The definition of the Wigner function is present and the method for computing the Wigner function is also described.

The Wigner function~\cite{Wigner1932, Hillery1984121, Kim_Wigner1990} for the  $n$-th energy state of the one DOF system is defined as
\begin{equation}
    \rho_{W_n}(x, p) = \dfrac{1}{2\pi \hslash}\int_{-\infty}^{\infty} \psi_n^{*}(x-\dfrac{\eta}{2}) \psi_n(x+\dfrac{\eta}{2}) \exp(\dfrac{i}{\hslash} p\eta)d\eta,
\label{defn::wigner1dof}    
\end{equation} 
where $\psi_n(x)$ is the $n$-th energy eigenfunction for the one DOF system and $\hslash$ is the reduced Planck constant. 
% $i$ is the imaginary unit, and $\eta$ is an integration variable. 
Even though the Wigner function for the harmonic oscillator has an analytical solution~\cite{Davies1975}, in general it can not be computed analytically due to the fact that the energy eigenfunction $\psi_n$ does not have an analytical solution. However, the integral~\eqref{defn::wigner1dof} can be computed numerically using the numerical solution of $\psi_n$. Thus, first we solve the time-independent Schr{\"o}dinger equation in $[a, b]$ using an uniform grid of $N$ points on the configuration coordinate, $x$
\begin{equation}
    x_j = a + \dfrac{(b - a)}{  (N - 1)} j, j = 0, 1, 2, ... N-1, 
\end{equation}
and consider a similar uniform grid of $N$ points in $[c, d]$ for the momentum coordinate, $p$
\begin{equation}
    p_k = c + \dfrac{d - c}{  (N - 1)} k, k = 0, 1, 2, ... N-1. 
\end{equation}
Using the uniform grid discretisation, the Wigner function at point $(x_j,p_k)$ becomes
\begin{equation}
\begin{aligned}
\left[\rho_{W_n}(x, p)\right]_{j,k} = \rho_{W_n}(x_{j}, p_{k}) 
&= \dfrac{1}{2\pi \hslash}\sum_{l} \psi_n(x_j-\dfrac{\eta_l}{2}) \psi_n(x_j+\dfrac{\eta_l}{2}) \exp(\dfrac{i}{\hslash} p_k\eta_l)\Delta\eta \\
&= \dfrac{1}{2\pi \hslash}\sum_{l} \psi_n(x_j-\dfrac{\eta_l}{2}) \psi_n(x_j+\dfrac{\eta_l}{2}) \cos{(\dfrac{p_k\eta_l}{\hslash})}\Delta\eta \\
&= \dfrac{1}{\pi \hslash}\sum_{l} \psi_n(x_j-\dfrac{\eta_l}{2}) \psi_n(x_j+\dfrac{\eta_l}{2}) \cos{(\dfrac{p_k\eta_l}{\hslash})}\Delta x.
\end{aligned}
\end{equation} 
We let $\Delta \eta = 2 \Delta x$ such that we can able to evaluate the $n$-th energy eigenfunction at position $x_j \pm \dfrac{\eta_l}{2}$ using the results we computed earlier. 

% \end{paracol}
\begin{figure}[!ht]
    % \widefigure
	\centering
	\subfigure[$\mathcal{D}=10.67, n=0$]{\includegraphics[width=0.32\textwidth]{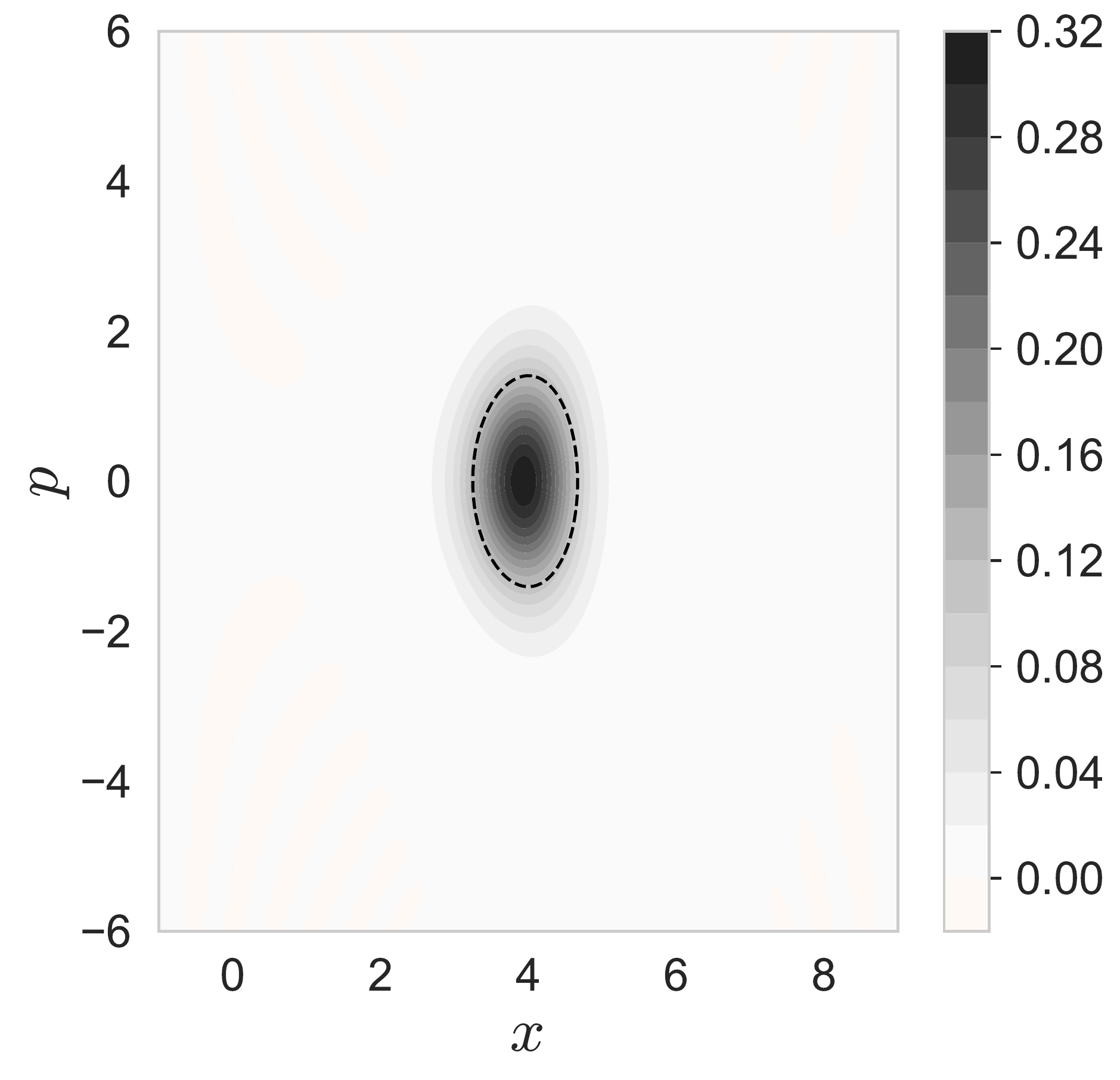}}
	\subfigure[$\mathcal{D}=10.67, n=1$]{\includegraphics[width=0.32\textwidth]{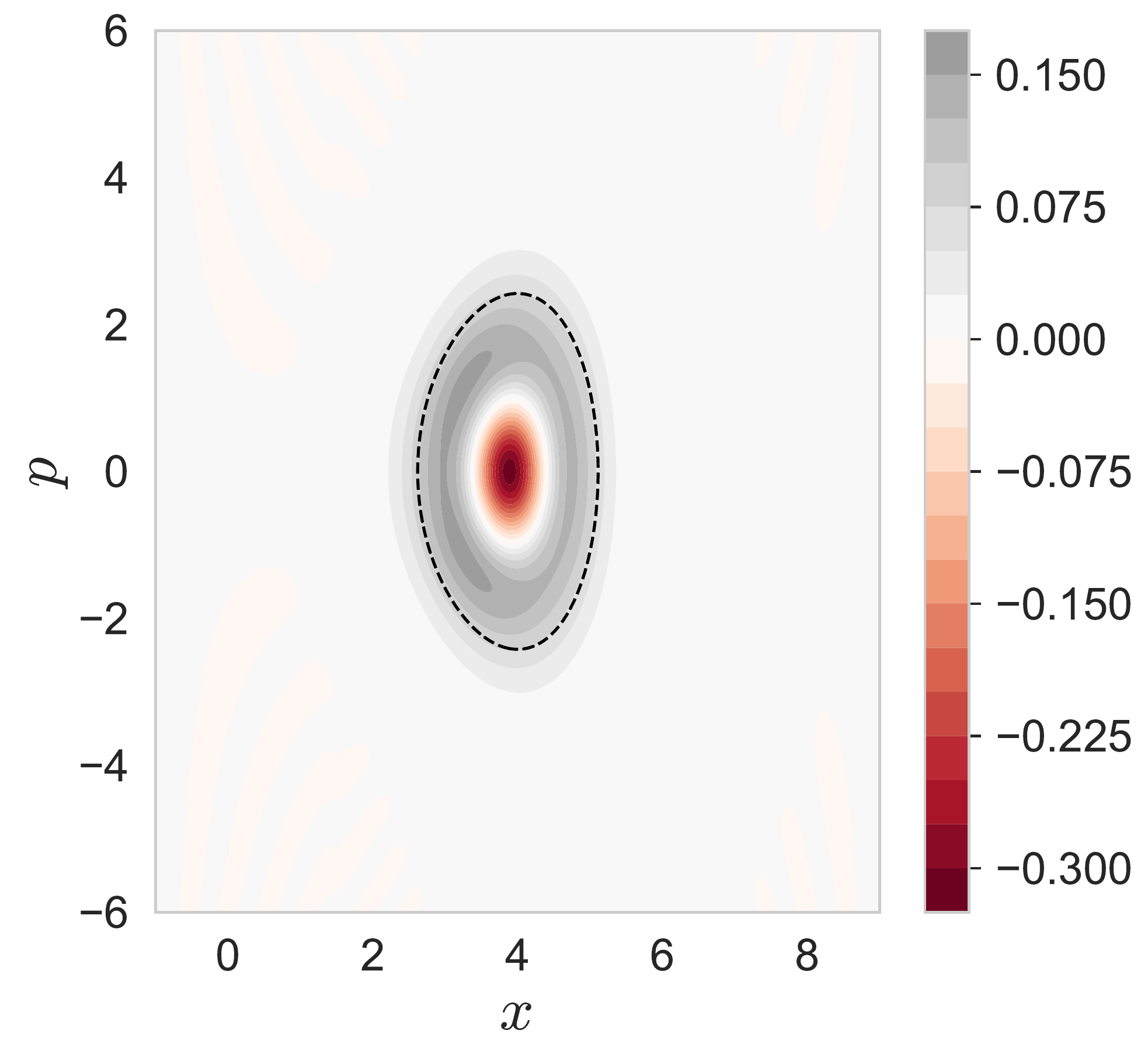}}
	\subfigure[$\mathcal{D}=10.67, n=4$]{\includegraphics[width=0.32\textwidth]{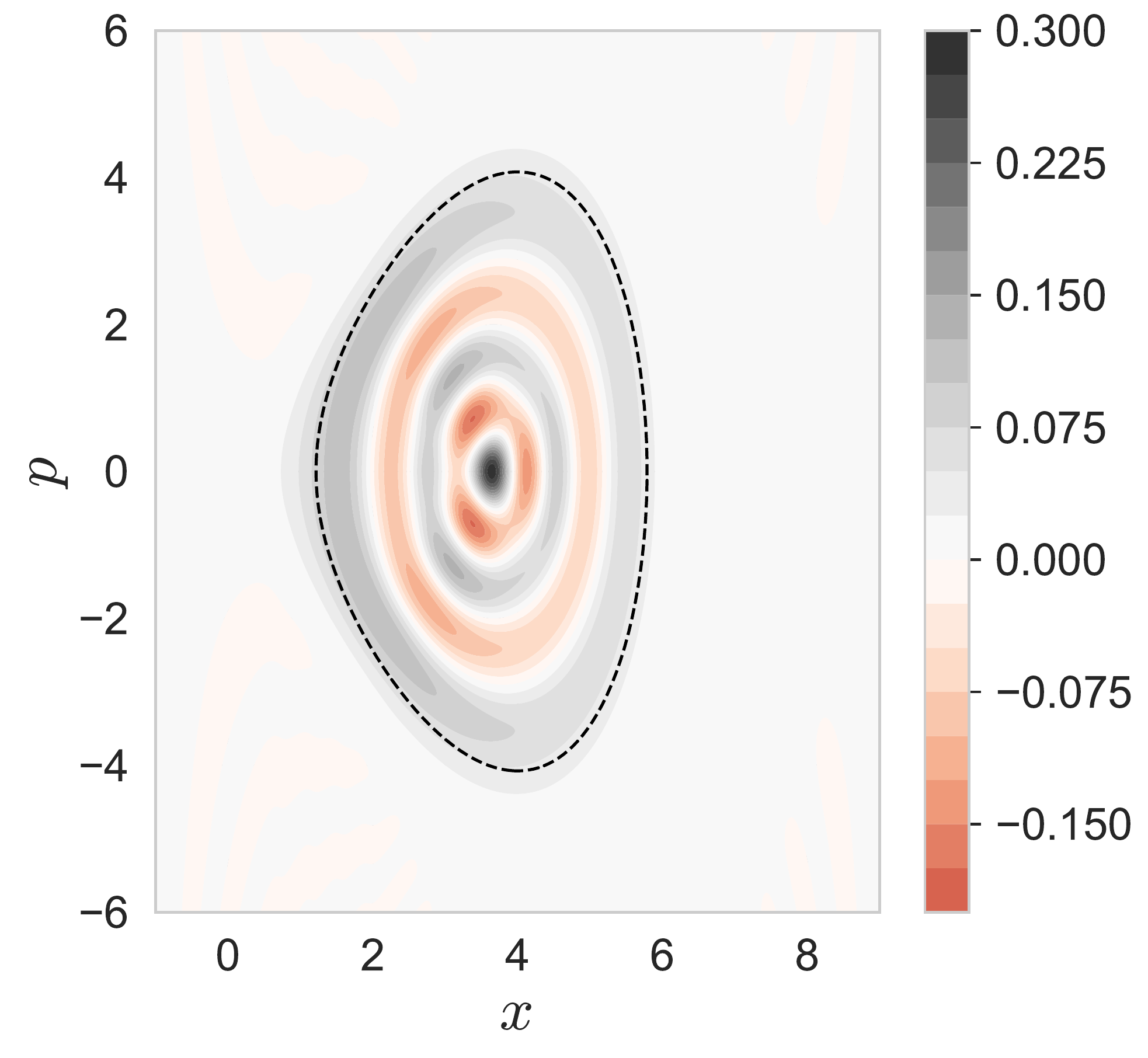}}
	\subfigure[$\mathcal{D}=2.67, n=0$]{\includegraphics[width=0.32\textwidth]{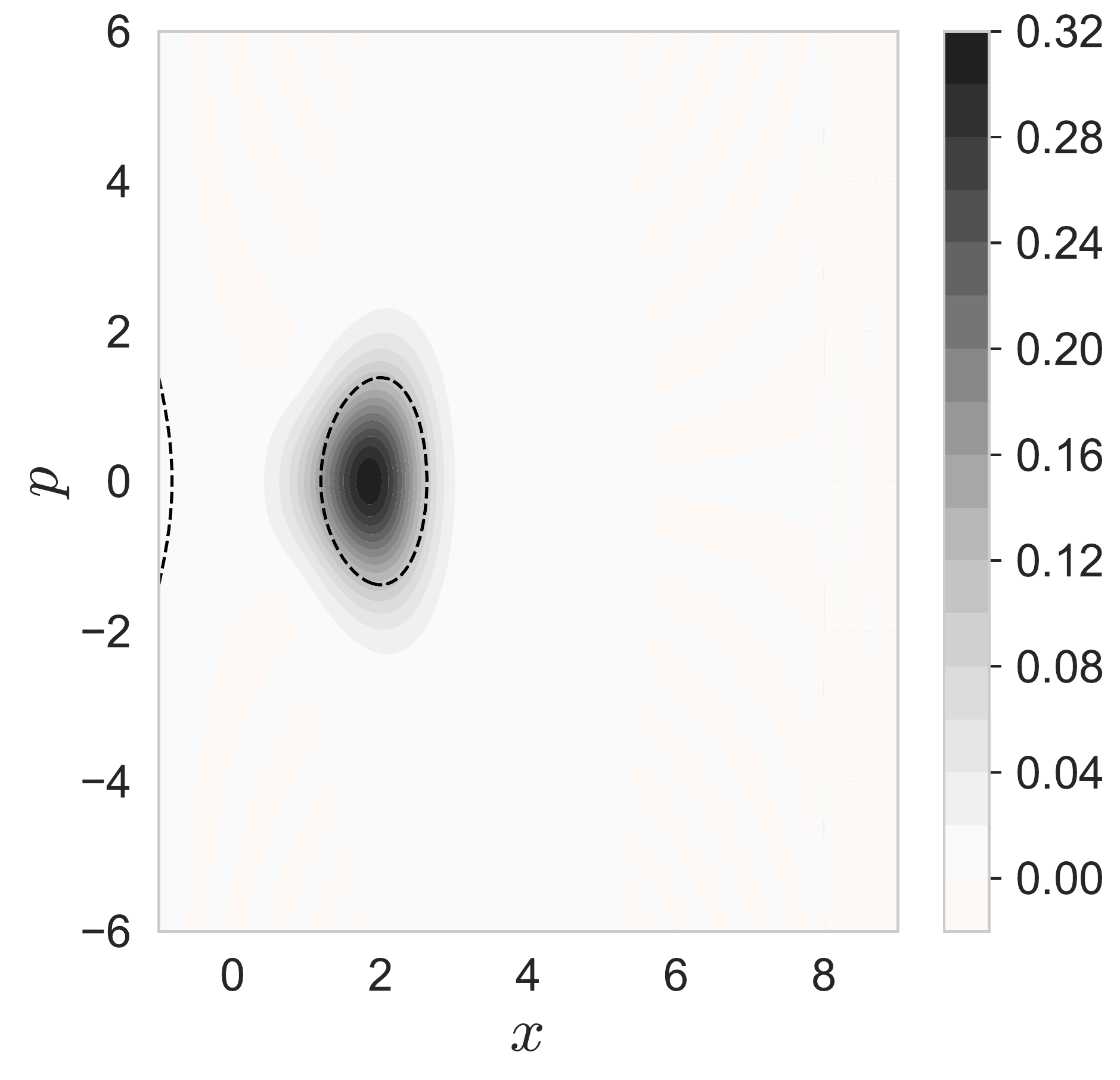}}
	\subfigure[$\mathcal{D}=2.67, n=1$]{\includegraphics[width=0.32\textwidth]{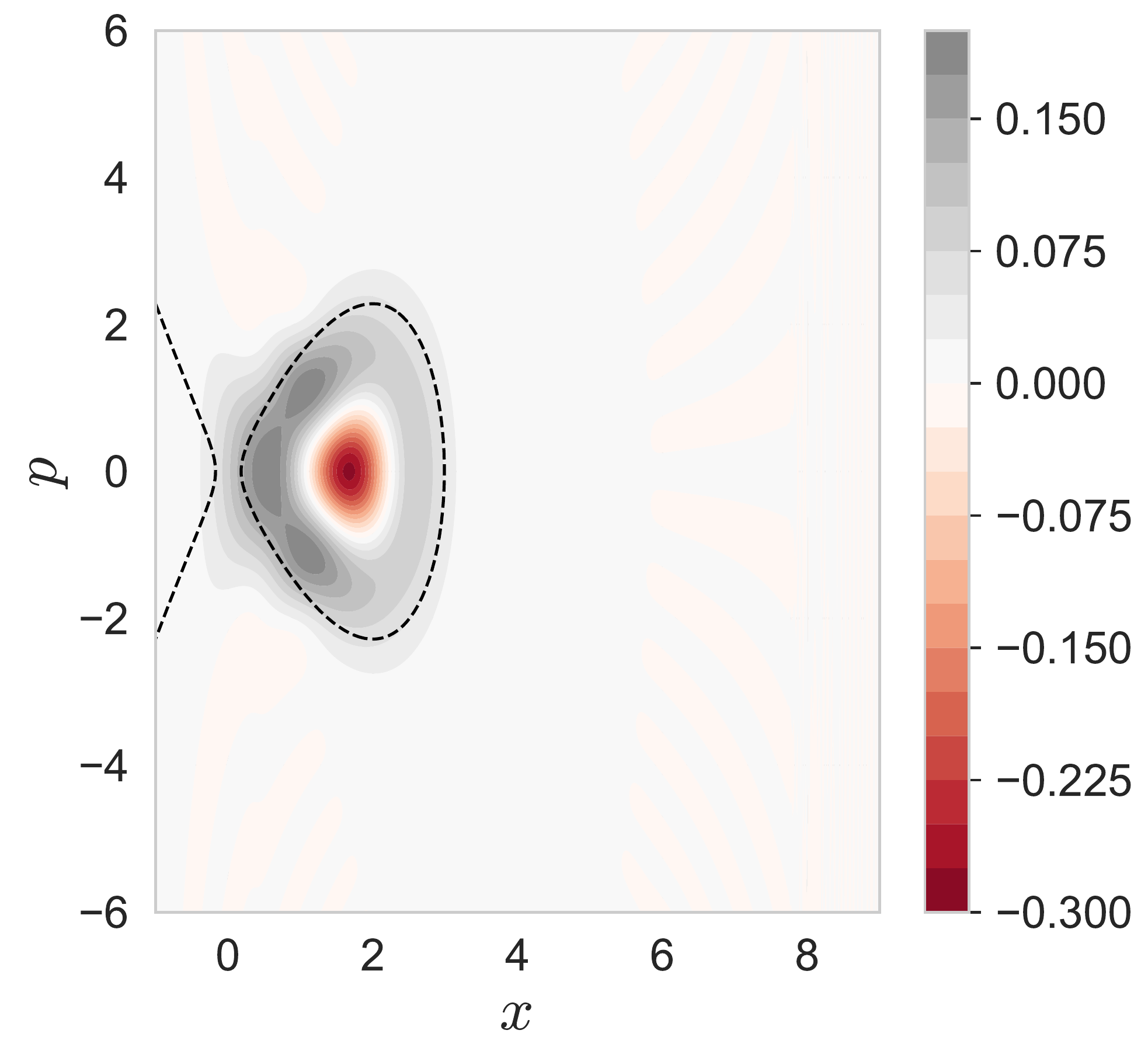}}
	\subfigure[$\mathcal{D}=2.67, n=4$]{\includegraphics[width=0.32\textwidth]{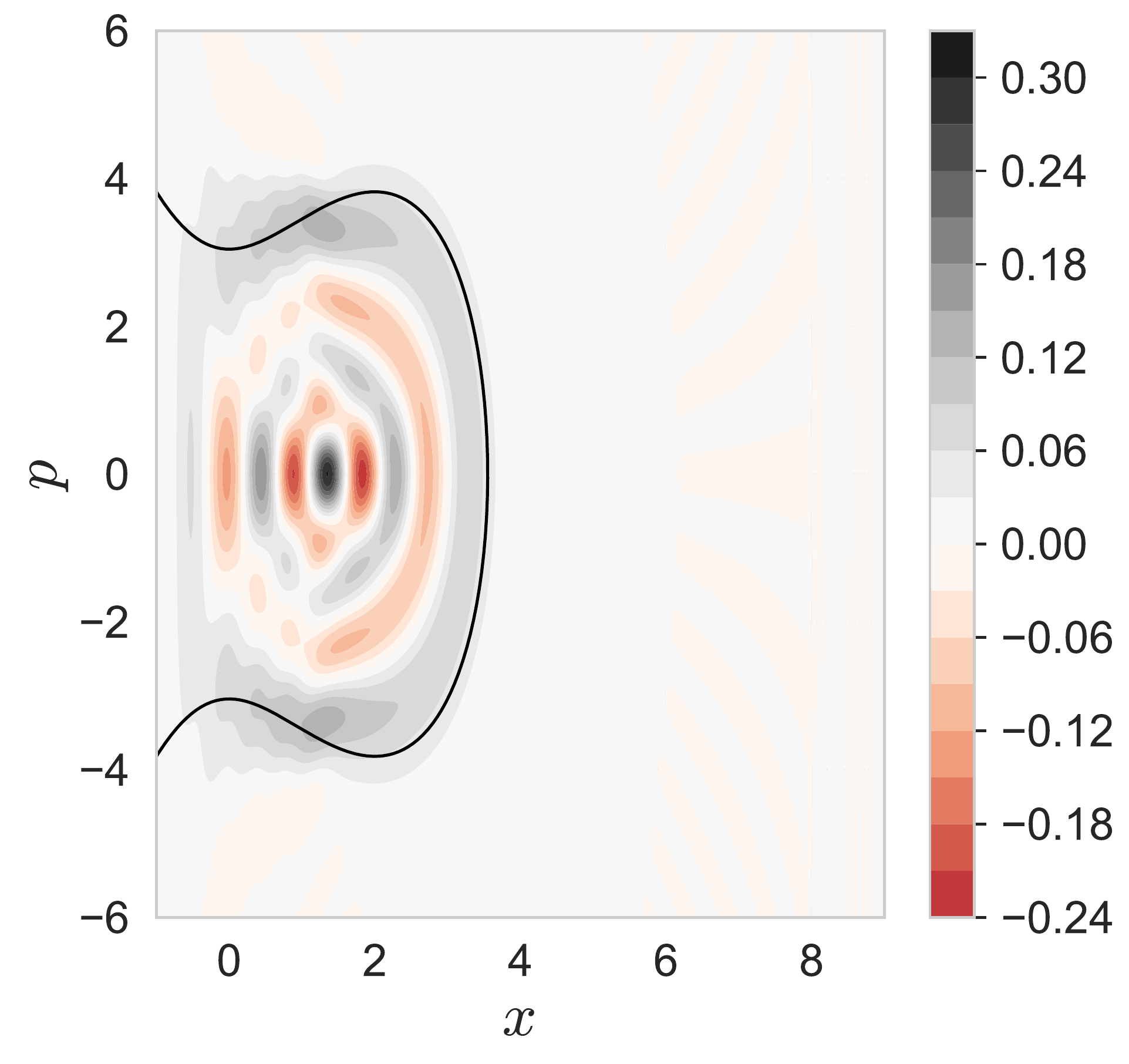}}
	\subfigure[$\mathcal{D}=0.43, n=0$]{\includegraphics[width=0.32\textwidth]{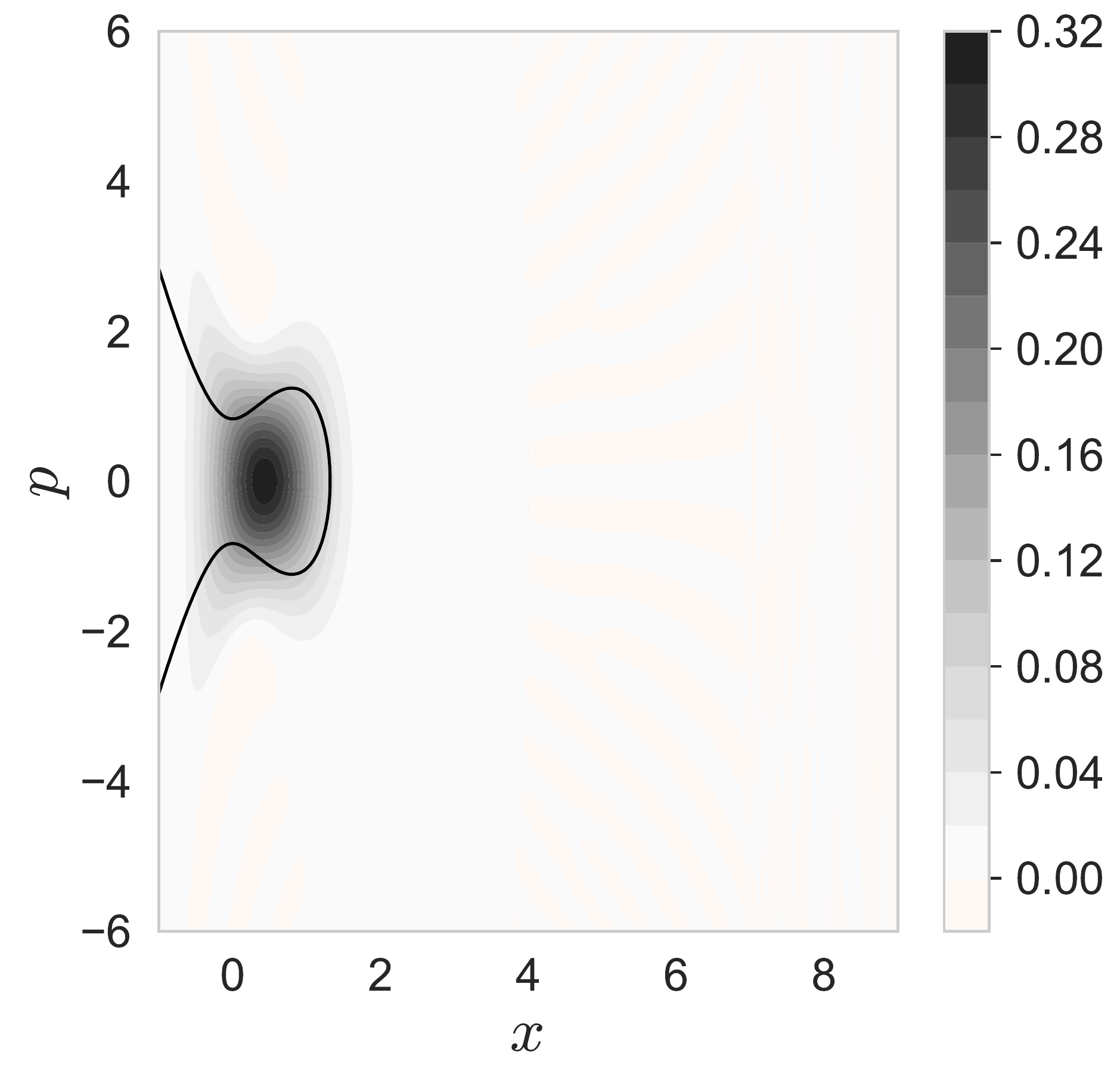}}
	\subfigure[$\mathcal{D}=0.43, n=1$]{\includegraphics[width=0.32\textwidth]{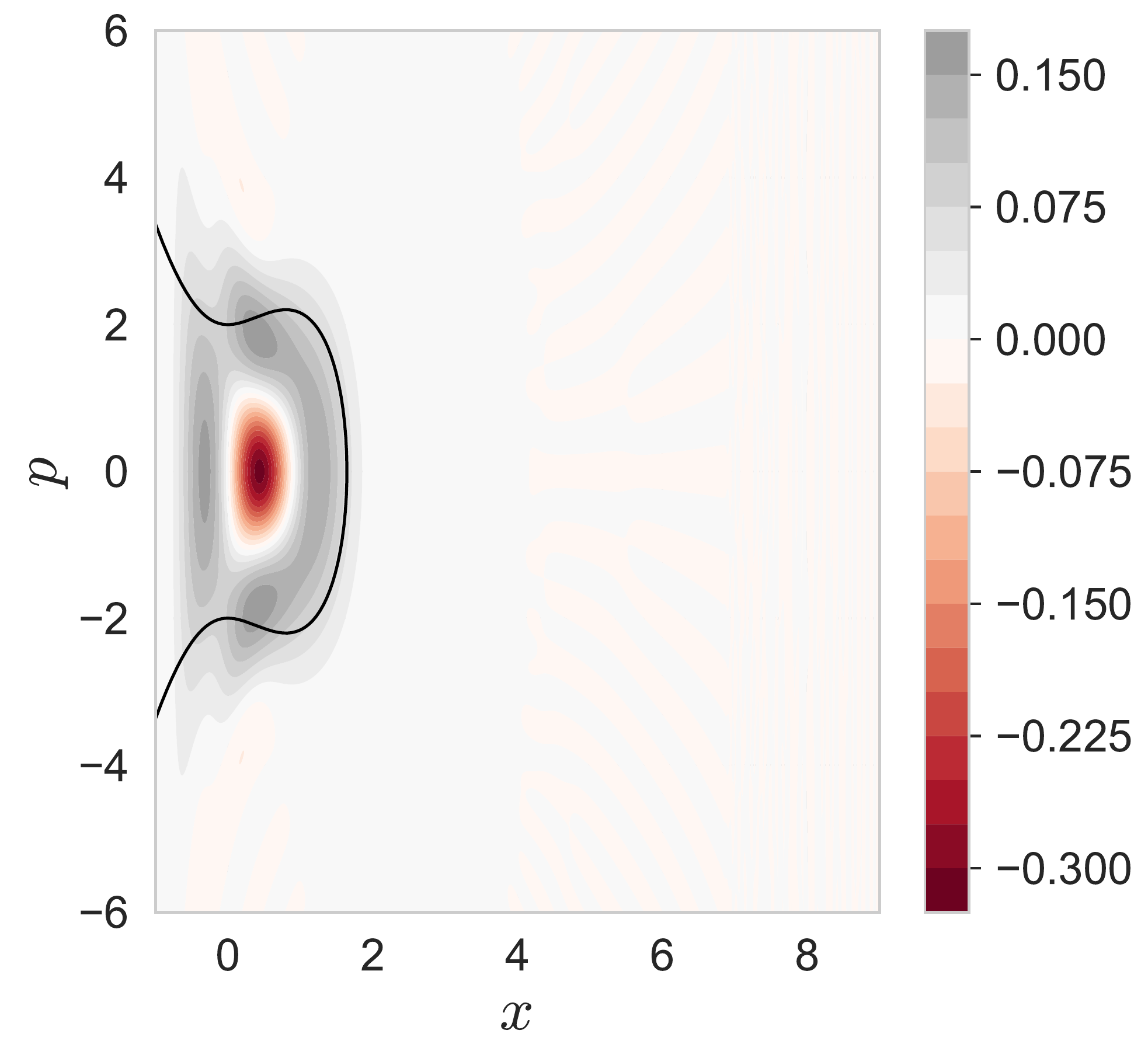}}
	\subfigure[$\mathcal{D}=0.43, n=4$]{\includegraphics[width=0.32\textwidth]{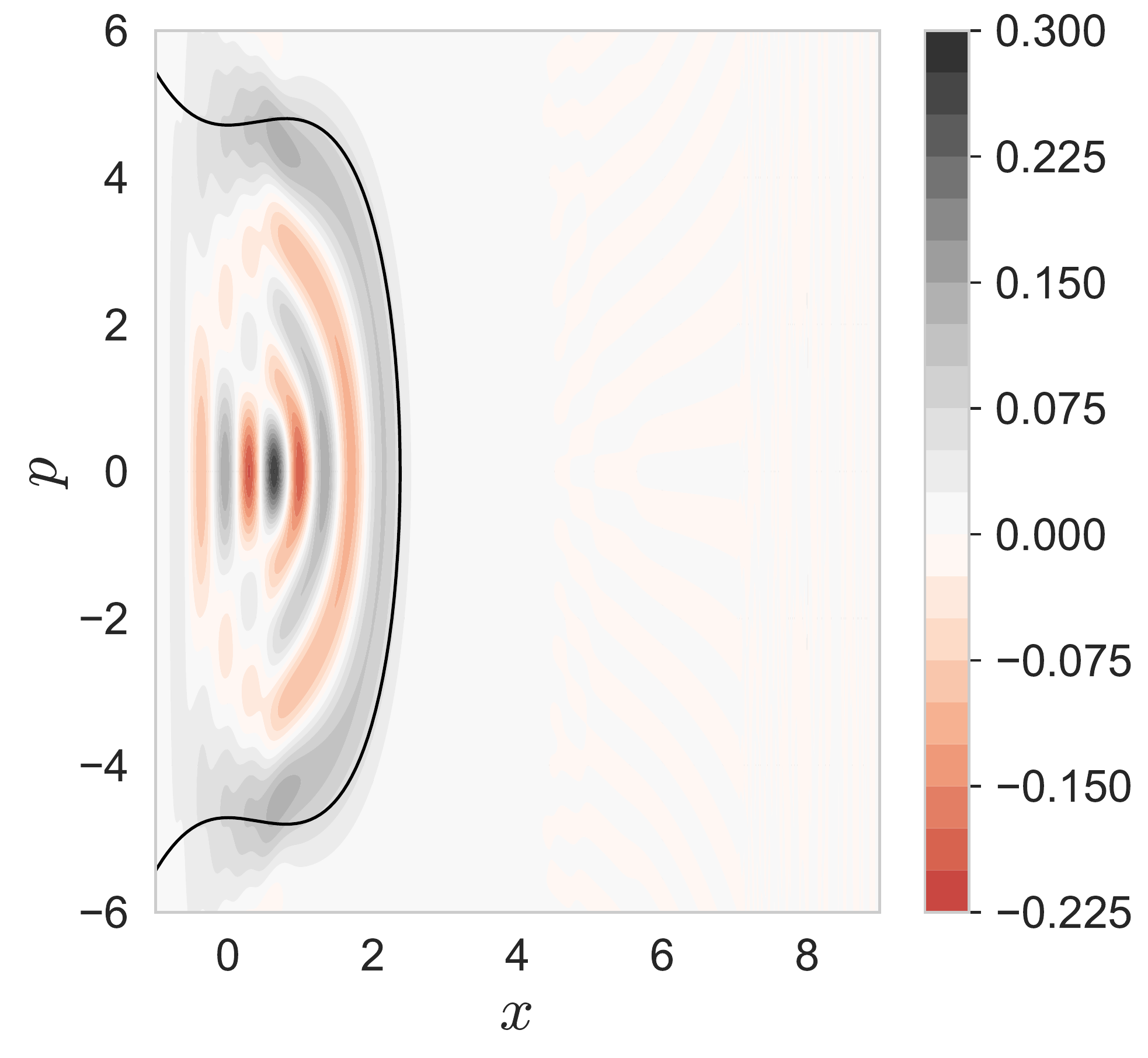}}
	\caption{Wigner function for $n = 0,1,4$ energy state. The coordinate space is chosen as $[-1,9]$, the momentum space is chosen as $[-6,6]$, number of grid points $N=599$, $m=1$, $\alpha=1, 2, 5 , \mu=4$ and $\hslash=1$. The black curve in each subfigure is the level curve of the classical Hamiltonian for total energy $e=E_n$.}
	\label{fig:quantum_sn1DOF_wigner}
\end{figure}
% \begin{paracol}{2}
% \linenumbers
% \switchcolumn

We recall that in classical dynamics the phase space trajectory with total energy $e$ is nonreactive when $e \leq 0$ and is reactive only when $e > 0$. The classical phase space trajectory is classified as reactive or nonreactive trajectory according to the total energy of the system relative to the energy of the saddle. Quantum mechanically, we describe the reactive and nonreactive behaviour in phase space by computing and analysing the Wigner function. We do this by calculating the probability of finding the system with classically nonreactive or reactive behaviour. The probability of finding the system with classically nonreactive behaviour is defined as the integral
\begin{equation}
    \int_{\mathcal{H} \leq 0} \rho_{W_n}(x, p)\ dx dp.
\end{equation}
We note that even though this integral is based on the Wigner function, it can be generalised to any other distribution functions defined on the phase space. 

\begin{figure}[!ht]
    % \widefigure
	\centering
	\subfigure[]{\includegraphics[width=0.32\textwidth]{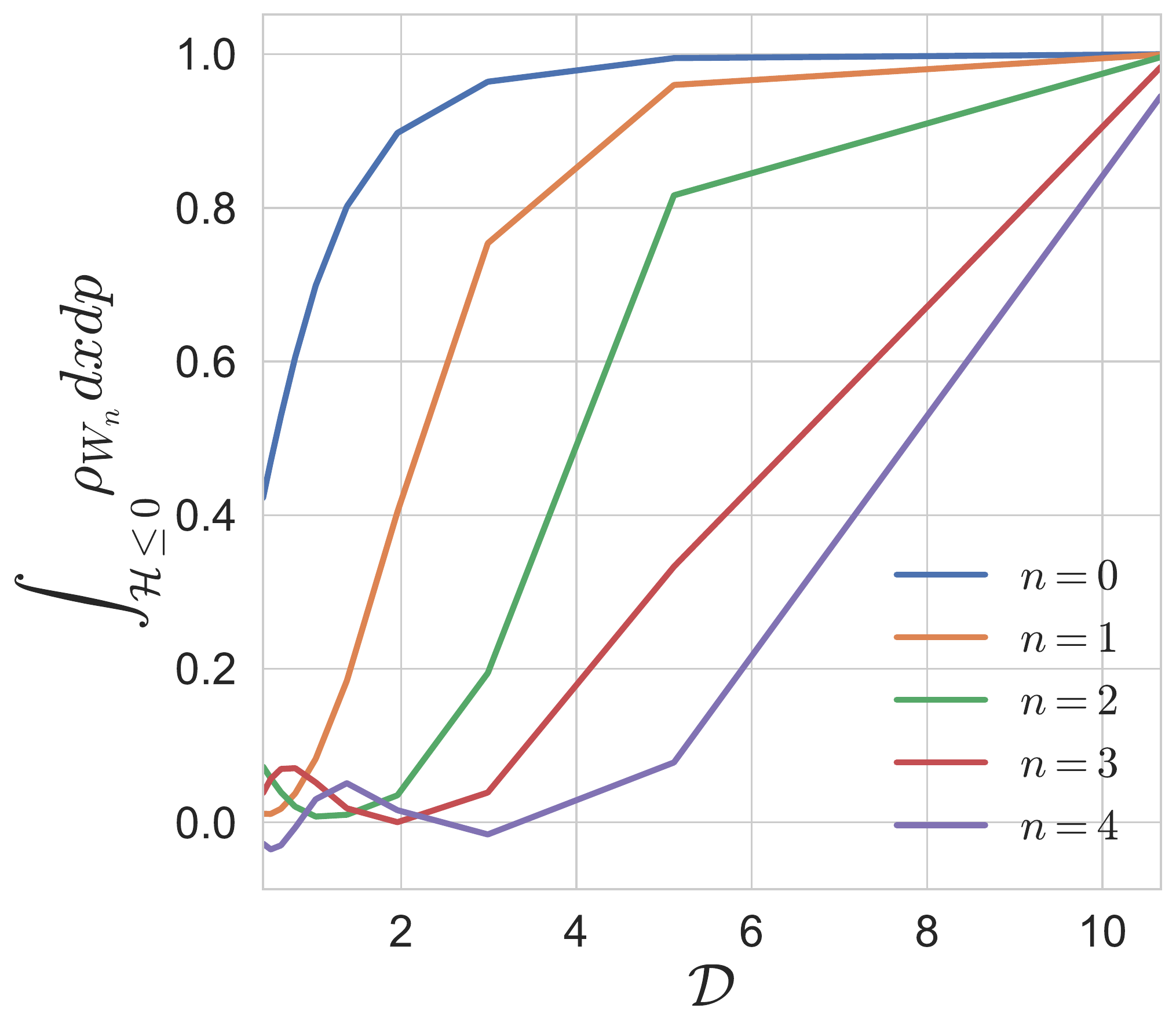}}
	%\subfigure[]{\includegraphics[width=0.32\textwidth]{figures_saddlenode1DOF/quantum_sd1DOF_husimi_integral_e_lesszero_dep.pdf}}
	\caption{Integral of the one DOF Wigner function on the domain $\mathcal{H} \leq 0$ for the first $5$ energy states. The coordinate space is chosen as $[-1,9]$, the momentum space is chosen as $[-6,6]$, number of grid points $N=599$, $m=1$, $\alpha=[1, 5] , \mu=4$ and $\hslash=1$.}
	\label{fig:quantum_sn1DOF_wigner_integral_e_lesszero_dep}
\end{figure}

We present the Wigner function for $n = 0,1,4$ energy state for the one DOF system with different values of $\mathcal{D}$ in Fig.~\ref{fig:quantum_sn1DOF_wigner} and the probability of finding the system with classically nonreactive behaviour in Fig.~\ref{fig:quantum_sn1DOF_wigner_integral_e_lesszero_dep}. In each subfigure of Fig.~\ref{fig:quantum_sn1DOF_wigner}, the phase space trajectory $\mathcal{H}=e=E_n$, where $E_n$ is the energy eigenvalue, is also plotted for comparison. The energy eigenvalue is obtained from the same quantum state used to compute the Wigner function. In Fig.~\ref{fig:quantum_sn1DOF_wigner} when the value of $\mathcal{D}$ is fixed, the non-zero values of the Wigner function lie on a larger domain in the phase space for higher energy states (as $n = 0,1,4$ correspond to Fig.~\ref{fig:quantum_sn1DOF_wigner} (a-c), respectively). Fig.~\ref{fig:quantum_sn1DOF_wigner_integral_e_lesszero_dep} shows that except for small values of $\mathcal{D}$, the probability of finding the system with classically nonreactive behaviour decreases when $n$ increases. When $\mathcal{D}$ is small, the probability of finding the system with classically nonreactive behaviour is close to $0$ except for $n=0$. This mismatches with what we observe in Fig.~\ref{fig:quantum_sn1DOF_wigner}(f). In this case, although the Wigner function has both positive and negative values in the region $\mathcal{H} \leq 0$, the probability of finding the system with classically nonreactive behaviour is close to zero. The probability of finding the system with classically nonreactive behaviour also shows consistent hill and a valley with upward rise for $n=2,3,4$ energy states with increasing well-depth. In Fig.~\ref{fig:quantum_sn1DOF_wigner} we observe as $\mathcal{D}$ increases, the Wigner function moves to the right in phase space which is consistent with what we observed for the mean position of the energy eigenfunction. Fig.~\ref{fig:quantum_sn1DOF_wigner_integral_e_lesszero_dep} shows that as the value of depth increases, the probability of finding the system with classically nonreactive behaviour increases and for large values of $\mathcal{D}$, the probability of finding the system with classically nonreactive behaviour is nearly one.

\section{Conclusions and outlook}
In this article, we studied the effect of the depth of the potential energy function on the quantum dynamics where the system's equilibrium points undergo saddle node bifurcation. We present the quantitative and qualitative analysis to show the connection between the well-depth and quantum mechanical quantities such as the energy eigenvalues, energy eigenfunctions, mean positions, position uncertainties of the position coordinate, and Wigner function. Quantitative analysis shows that for any excited state we considered, as the value of $\mathcal{D}$ increases, the energy eigenvalue decreases linearly, the mean position moves away from the saddle, and the position uncertainty becomes constant. We also observe from Figure.~\ref{fig:quantum_sn1DOF_wigner_integral_e_lesszero_dep} that the probability of finding the system with classically nonreactive behaviour increases and approaches one whereas classically this probability is either zero (when the total energy $e$ is less than zero) or one (when the total energy $e$ is greater than zero).

Future work will be to use the formulation of the well-depth to study its influence on the quantum dynamics of a two degree-of-freedom and time-dependent one degree-of-freedom system. We will study how the loss of integrability (chaos) to obtain the classical dynamics for some parameter values affect the quantum dynamics.

% In this case the classical Hamiltonian needs not to be integrable and can have complicated classical dynamics such as chaos in phase space. 
% It is interesting to see the effect of the depth of the potential energy function on such systems and how it relates to reaction dynamics. 
%%%%%%%%%%%%%%%%%%%%%%%%%%%%%%%%%%%%%%%%%%
\vspace{6pt} 


\bibliographystyle{abbrv}
\bibliography{references}

\appendix
\section{Methods for computing the quantum states}

\subsection{The finite difference(FD) method}
The idea of the FD method is to discretise and solve the time-independent Schr\"{o}dinger equation in Eqn.~\eqref{eqn:time_inde_Schrodinger_1d} as an eigenvalue problem in configuration space. The time-independent Schr\"{o}dinger in Eqn.~\eqref{eqn:time_inde_Schrodinger_1d} can be written as
\begin{equation}
    \left(-\dfrac{\hslash^2}{2} \frac{\partial^2}{\partial x^2} + V(x) \right) \psi(x) = E \psi(x)
\end{equation}
As the coordinate $x$ is in a continuous range, we restrict the range of $x$ into a finite interval $[a,b]$ and discretise the interval into $N$ equally spaced grid points with spacing $\Delta x$ such that
\begin{equation}
    x \rightarrow a + j \Delta x,\quad j = 0, ..., N-1
\end{equation}
Note that $ (N-1)\Delta x = b-a$. The discrete analogue of the first derivative of $\psi(x)$ is given by:
\begin{equation}
    \frac{\partial \psi(x)}{\partial x} = \dfrac{\psi(x + \Delta x) - \psi(x )}{\Delta x}
\end{equation}
Applying a similar first derivative formula to $\partial \psi(x)/\partial x$, we get the discrete analogue of the second derivative of $\psi(x)$:
\begin{equation}
    \frac{\partial^2 \psi(x)}{\partial x^2} =\dfrac{\frac{\partial \psi(x)}{\partial x} - \frac{\partial \psi(x-\Delta x)}{\partial x}}{\Delta x} =\dfrac{\psi(x + \Delta x) - 2\psi(x) + \psi(x - \Delta x)}{(\Delta x)^2}
\end{equation}
The discretisation result of $\hat{\mathcal{H}}$ is therefore a $N \times N$ Hamiltonian matrix $\tilde{\mathcal{H}}$ and we solve it with dirichlet boundary condition $\psi(a) = \psi(b) =0$. The boundary condition means we can treat $\tilde{\mathcal{H}}$ as a $(N-2) \times (N-2)$ matrix and we only need to compute the matrix elements for the remaining $N-2$ grid points:
% \end{paracol}
% \nointerlineskip
\begin{equation}
\begin{aligned}
\tilde{\mathcal{H}} = \tilde{\mathcal{T}} + \tilde{\mathcal{V}} = &
-\dfrac{\hslash^2}{2(\Delta x)^2}\begin{pmatrix}
-2 &1 & 0&... &...& 0\\
1 & -2 &  1& 0 &...& \vdots\\
0& 1 &  \ddots & \ddots & \ddots & \vdots\\
\vdots& \ddots &\ddots  & \ddots & \ddots & 0\\
0 & \dots & 0  & 1 & -2 &1\\
0 & \dots & \dots  & 0 & 1 &-2\\
\end{pmatrix}
+ \\
& \begin{pmatrix}
V(a + \Delta x) &0 & \dots&... &...& 0\\
0 & V(a + 2\Delta x) &  0 & \dots &...& \vdots\\
\vdots& 0 &  \ddots & \ddots & \ddots & \vdots\\
\vdots& \ddots &\ddots  & \ddots & \ddots & \vdots\\
\vdots & \dots & \dots  & 0 & \ddots &0\\
0 & \dots & \dots  & \dots & 0 &V(a + (N-2)\Delta x)\\
\end{pmatrix}
\end{aligned}
\end{equation}
% \begin{paracol}{2}
% \linenumbers
% \switchcolumn
Thus solving the time-independent Schr\"{o}dinger equation Eqn.~\eqref{eqn:time_inde_Schrodinger_1d}
is equivalent to finding the eigenvalues $E_n$ and the normalised eigenvectors $\psi_{n,k}$ of the matrix $\tilde{\mathcal{H}}_{jk}$ where the normalisation condition becomes $\sum_k \psi_{n,k}^{*}\psi_{n,k} = 1/(\Delta x)$.
\begin{equation}
\sum_{k}\tilde{\mathcal{H}}_{jk} \psi_{n,k}(x) = E_n \psi_{n,k}(x)
\end{equation}

\subsection{Perturbation type solution}
As we are interested in the dynamics near the centre equilibrium point, we Taylor expand the potential energy function $V(x)$ near $x_2^e$ and get
\begin{align}
    V(x) &= V(x_2^e) + \dfrac{1}{2}V''(x_2^e)(x-x_2^e)^2 + \dfrac{1}{6}V'''(x_2^e)(x-x_2^e)^3\\
    &= - \dfrac{4\mu^{3/2}}{3\alpha^2} + \dfrac{1}{2}2 \sqrt{\mu} (x-x_2^e)^2 + \dfrac{1}{6} 2\alpha (x-x_2^e)^3. \label{eqn::quantum_pes_harm+perturb}
\end{align}
This shows that the system near $x_2^e=2\sqrt{\mu}/\alpha$ is a quantum harmonic oscillator with a cubic perturbation when $\alpha$ is small.

Using Eqn.~\eqref{eqn::quantum_pes_harm+perturb}, we can see that the energy eigenvalue of the newly defined energy eigenstate is approximately (ignoring the cubic perturbation term) 
\begin{equation}
    - \dfrac{4\mu^{3/2}}{3\alpha^2} + \hslash \sqrt{2\sqrt{\mu}} (n + \dfrac{1}{2}) = - \mathcal{D} + \hslash \sqrt{2\sqrt{\mu}} (n + \dfrac{1}{2}), \quad n = 0, 1, 2, ...
\end{equation}

\end{document}